\documentclass[twocolumn, aps, 10pt, pra]{revtex4-1}

\usepackage{mathbbol}              % for math symbols.

\usepackage{graphicx}

\usepackage{ulem}

\usepackage{color}
\usepackage{hyperref}
\usepackage[usenames,dvipsnames]{xcolor}
\usepackage{amsmath, amsthm, amssymb}

\begin{document}

\title{Microwave control of the interaction between two optical photons}

\date{\today}
\author{D. Maxwell}
\altaffiliation[Now at: ]{Department of Physics,
Vivian Tower,
Swansea University,
Singleton Park,
Swansea SA2 8PP,
UK.}
\email{daniel.thomas.maxwell@cern.ch}
\author{D. J. Szwer}
\author{D. Paredes-Barato}
\author{H. Busche}
\author{J. D. Pritchard}
\altaffiliation[Now at: ]{Department of Physics,
University of Wisconsin,
1150 University Avenue,
Madison WI 53706-1390, USA.}
\author{A. Gauguet}
\altaffiliation[Now at: ]{Laboratoire Collisions Agr\'egats R\'eactivit\'e,
Universit\'e Paul Sabatier,
B\^at.\ 3R1b4,
118 route de Narbonne,
31062 Toulouse Cedex 09,
France.}
\author{M. P. A. Jones}
\author{C. S. Adams}
\email{c.s.adams@durham.ac.uk}
\affiliation{Joint Quantum Centre (JQC) Durham-Newcastle,
Department of Physics,
Durham University,
Rochester Building,
South Road,
Durham DH1 3LE,
England.}

\begin{abstract}  
A microwave field is used to control the interaction between pairs of optical photons stored in highly excited collective states (Rydberg polaritons). We show that strong dipole-dipole interactions induced by the microwave field destroy the coherence of polariton modes with more than one Rydberg excitation. Consequently single-polariton modes, which correspond to single stored photons, are preferentially retrieved from the sample. Measurements of the photon statistics of the retrieved light field also reveal non-trivial propagation dynamics of the interacting polaritons. 
\end{abstract}

\maketitle

The ability to perform quantum computations with photons is limited by weak photon-photon interactions.  Consequently, research has focused either on cavity enhancement of the non-linearity in both the optical \cite{Birn2005} and microwave domains \cite{Kirchmair2013}, or linear optics quantum computing (LOQC), where the nonlinearity comes from the projective (and hence probabilistic) measurement of part of the system \cite{Knill2001}.

% Recently, it has been shown that large single-photon non-linearities are also accessible without a cavity by coupling optical photons to highly-excited Rydberg states \cite{Dudin2012, Peyronel2012, Maxwell2013,Hof2013,Baur2013,Li2013}.  The potential of this emerging field of Rydberg quantum optics is particularly promising because the long-range character of the dipole-dipole interaction between Rydberg atoms makes it possible to achieve a non-local optical non-linearity \cite{Fri2005,Sevincli2011}. For example, a non-local interaction can circumvent a fundamental limitation of the local Kerr-like optical non-linearity for all-optical information processing \cite{Shapiro2006, Banacloche2010}.  In addition, it is possible to modify the atom-light interaction \cite{Mohapatra2008} and the interactions between Rydberg states using external fields \cite{Tanasittikosol2011, Brekke2012,Bariani2012,SaffWalk}. By interfacing Rydberg quantum optics with microwave control of the dipole-dipole interaction, it is possible to realise a fully deterministic photonic phase gate \cite{Paredes-Barato2013}.

Recently, it has been shown that large single-photon non-linearities are also accessible without a cavity by coupling optical photons to highly-excited Rydberg states \cite{Dudin2012, Peyronel2012, Maxwell2013, Hof2013, Baur2013, Li2013}.  The potential of this emerging field of Rydberg quantum optics is particularly promising because the long-range character of the dipole-dipole interaction between Rydberg atoms makes it possible to achieve a non-local optical non-linearity \cite{Fri2005, Sevincli2011}. Such a non-local interaction can circumvent a fundamental limitation of the local Kerr-like optical non-linearity for all-optical information processing \cite{Shapiro2006, Banacloche2010}.  A promising route to all-optical photonic processing makes use of a microwave field to control the photon-photon interaction \cite{Paredes-Barato2013}.  In this paper we demonstrate the basis of this route by showing that an external microwave field can modify the photon statistics of an optical probe beam.  The microwave control operates by modifying the long range dipole-dipole interaction between photons stored as Rydberg polaritons.  In the current experiment this is manifest as a dephasing of the polariton spin wave \cite{Bariani2012} leading to photon loss.  By changing the geometry this could be adapted to realise a fully deterministic photonic phase gate \cite{Paredes-Barato2013}.

\begin{figure}[!bt]
\includegraphics[width=\columnwidth]{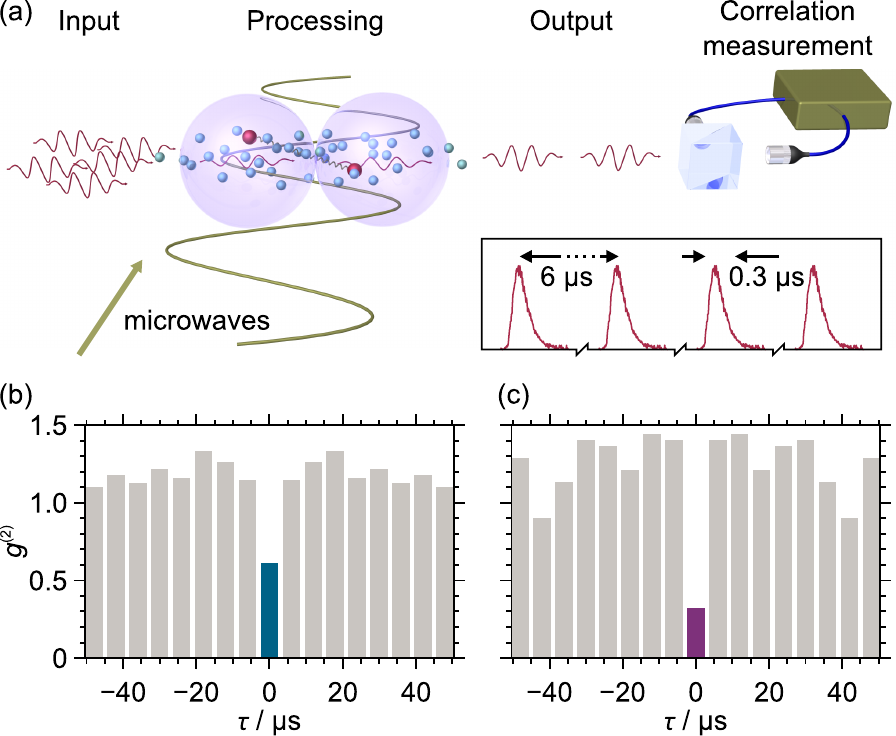}
\caption{(Color online)  Control of optical photon statistics by an external microwave field. (a) Optical signal photons are stored in a cold atomic cloud as Rydberg polaritons. Subsequently, an external microwave field controls the state and the interaction between neighboring polaritons. Finally the optical field is retrieved and the second order correlation function, $g^{(2)}$, is analysed using two single-photon counters. The $6~\mu$s separation between the retrieved pulses, shown in the inset, corresponds to the repetition rate of the experiment. (b) The normalised $g^{(2)}$ function of the retrieved signal as a function of time delay, $\tau$, between the two detectors. No microwave field is applied during the storage interval in this case. The suppression of $g^{(2)}$ at $\tau=0$ is a consequence of dipole blockade during the photon storage process. (c) The application of a microwave field during the storage interval results in significantly enhanced suppression of $g^{(2)}$ at $\tau=0$. Note that the entire retrieved pulse has been binned for both sets of $g^{(2)}$ measurements, and the $g^{(2)}$ functions are background corrected using the measured signal to noise ratio \cite{Orrit2002}.}
\label{Apparatus}
\end{figure}

% In this paper we demonstrate microwave control of the interaction between two optical photons stored in a cold atomic gas. The experiments build upon the work described in reference \cite{Maxwell2013}.

The work described in this paper builds on results published in reference \cite{Maxwell2013}, where it was shown that a microwave field can be used to control both the quantum state and the interaction-induced dephasing of Rydberg polaritons.  Here, through an improvement in experimental techniques that lead to increased signal photon rates, we directly show that the photon pair correlations are modified by the application of a microwave pulse.  This provides experimental confirmation of theoretical proposals based on the dephasing of light-induced atomic excitations \cite{Bariani2012, Stan2012, Joner2011}.  Our results also exhibit evidence of a photon number dependent group delay, which could potentially be used as a filtering device for Fock states \cite{Niko2010,Suz2011}.

We use electromagnetically induced transparency (EIT) \cite{Mohapatra2008, Tanasittikosol2011} to store optical photons in Rydberg states, which have a large principal quantum number \cite{Gala}. Under conditions of EIT, a medium is made transparent to an optical signal field which would otherwise be resonantly scattered \cite{Fleischhauer2005}. This is achieved by coupling the signal field to a third, long-lived atomic level with a control field. In our system this long-lived atomic level is a Rydberg state. The signal field creates collective Rydberg excitations, known as Rydberg polaritons, which are quasi-particles containing a mixture of photonic and excitonic character \cite{Fleischhauer2000}. The mixing between the two components is determined by the control field intensity. During storage where the control field intensity approaches zero, the polaritons are mainly atomic in nature and thus the signal photons are stored in the long-lived Rydberg state. Under ideal conditions the phase imprinted into the atomic medium during the storage process means that, upon increasing the control field intensity, the retrieved field is emitted into a well-defined mode, even when there is only one stored photon \cite{Scully2006, Scully2010}.

%The signal photons can be stored in this long-lived state, and then be subsequently retrieved from the medium since their group velocity is dependent upon the intensity of the control field \cite{Dudin2013}. Under ideal conditions the phase imprinted into the atomic medium during the storage process means that the retrieved field is emitted into a well-defined mode. 

The experimental process is shown schematically in figure \ref{Apparatus}(a). Full details can be found in reference \cite{Maxwell2013}. A signal pulse containing approximately 30 photons is incident upon a laser-cooled cloud of $^{87}\mathrm{Rb}$ atoms. Electromagnetically induced transparency is performed with signal light at 780.2~nm\ (addressing the $5s~^2S_{1/2}(F=2) \rightarrow 5p~^2P_{3/2}(F=3)$ transition), and control light at 479.8~nm\ (addressing the $5p~^2P_{3/2}(F=3) \rightarrow 60s~^2S_{1/2}$ transition). The signal photons are stored in the $60s~^2S_{1/2}$ Rydberg state for a period of approximately 900~ns by lowering the intensity of the control field. During the storage interval a microwave pulse of 150~ns duration resonantly couples the $60s~^2S_{1/2}$ state to the $59p~^2P_{3/2}$ state. The retrieved light field is detected by a pair of single-photon detectors allowing measurements of the second-order correlation function, $g^{(2)}$. In addition to correlation measurements, photon counting provides two distinct methods of observing photon-photon interactions induced by the microwave field. Each of the methods is discussed in turn below.

First let us consider the storage process alone where no microwave field is applied to the sample. The strong van der Waals dipole-dipole interactions between Rydberg atoms lead to the phenomenon of dipole blockade \cite{Lukin2001}, where only a single Rydberg excitation is allowed within a region known as the blockade sphere. Each Rydberg excitation is delocalised over all the atoms in the corresponding blockade sphere. In our system the radius of the blockaded region, ${R_{\mathrm{o}}}$, is determined by the EIT linewidth, $\Delta_{\mathrm{EIT}}$. The blockade radius is given by $R_{\mathrm{o}}=(C_{6}/\hbar\Delta_{\mathrm{EIT}})^{1/6}$, where $C_{6}$ is the van der Waals coefficient. For the $60s~^2S_{1/2}$ Rydberg state used in the experiments described in this paper, and for $\Delta_{\mathrm{EIT}}/2\pi=1$~MHz, ${R_{\mathrm{o}}}\approx 7~\mu$m. The effect of dipole blockade is to spatially localise each Rydberg polariton over the correlation length $R_{\mathrm{o}}$, as illustrated in figure \ref{Apparatus}(a). Due to the finite size of our atomic sample, the blockade radius ${R_{\mathrm{o}}}$ determines the maximum number of polaritons that can be written into the medium. Since the width of the medium is smaller than ${R_{\mathrm{o}}}$, the system is effectively one-dimensional. Figure \ref{Apparatus}(b) shows a measurement of the photon statistics of the retrieved field, giving $g^{(2)}(0)=0.62 \pm 0.06$. The anti-bunching of the retrieved signal is a signature of dipole blockade. In addition to dipole blockade, dephasing of the polaritons due to short-ranged van der Waals interactions has been shown to play a role in their spatial correlations \cite{Dud2012}.

Longer range interactions between the polaritons can be induced by coupling them to neighboring Rydberg states with opposite parity. The microwave field which is applied during the storage interval causes the polaritons to interact via longer range resonant dipole-dipole interactions. This introduces a second blockade scale \cite{Tanasittikosol2011,Brekke2012}, denoted ${R}_{\mu}$, that is dependent on the Rabi frequency of the microwaves, $\Omega_{\mu}$, i.e. $R_{\mu}=(C_{3}/\hbar\Omega_{\mu})^{1/3}$, where $C_{3}$ is the resonant dipole-dipole interaction coefficient. If two polaritons are separated by a distance comparable to ${R}_{\mu}$, then the strong dipole-dipole interactions result in dephasing of the polaritons \cite{Bariani2012} through resonant energy exchange \cite{Ates2008,Ditz2008}. This destroys the directionality of the read-out and therefore suppresses the retrieved photon signal. 

\begin{figure*}[!bt]
\includegraphics{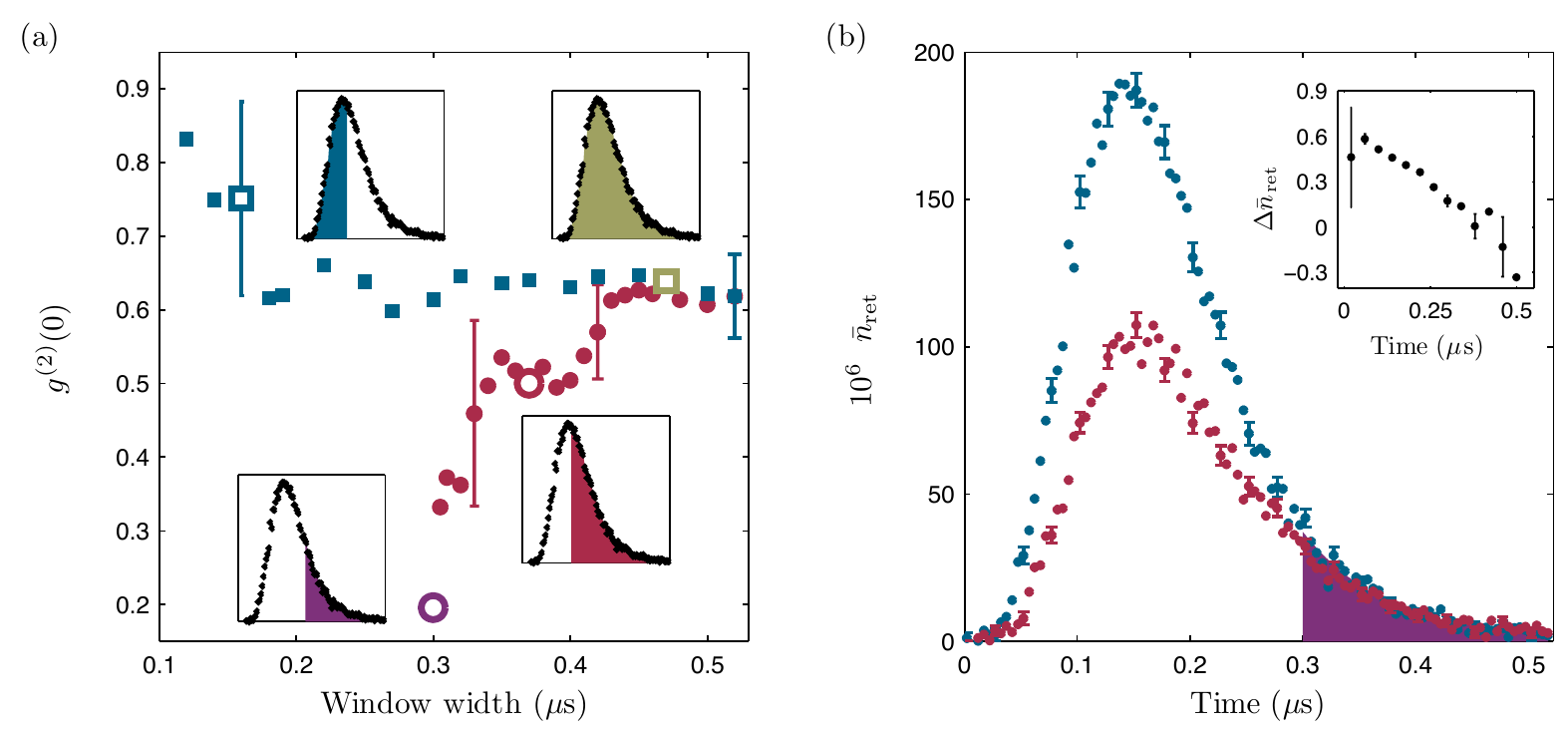}
\caption{(Color online)  Photon number dependent group delay. (a) The variation of $g^{(2)}(0)$ depending on the sampling window of the retrieved pulse. The width of the sampling window is varied, fixing one end at the trailing edge (red circles) or the leading edge (blue squares) of the retrieved pulse. The insets show the retrieved pulse shape with the corresponding sampling window highlighted. Each inset corresponds to the adjacent hollowed data point. No microwave coupling has been applied in this case. The variation of $g^{(2)}(0)$ suggests that the group delay of the retrieved pulse depends on the photon number. (b) The retrieved pulse shapes for the case where no microwave coupling is applied (blue, upper curve), and after microwave coupling (red, lower curve). The amplitude of the pulses are scaled in terms of the average number of photons retrieved per store-and-retrieve cycle, $\bar{n}_{\mathrm{ret}}$. After microwave coupling there is a suppression in the amplitude of the retrieved field due to dephasing of multi-polariton states. However, there is a large degree of overlap of the pulses in their trailing edges (purple area). This corresponds to the region where the single photon mode (corresponding to a single polariton) should dominate. The overlap in the trailing edges is shown more clearly in the inset. Here $\Delta\bar{n}_{\mathrm{ret}}$ is the difference between the two retrieved pulses, normalised by the case of no microwave coupling. A larger (40~ns) bin width has been used to calculate the normalised pulse difference in order to improve the counting statistics.}
\label{fockfil}
\end{figure*}

It has been proposed that the effects of dephasing should also be manifest in the photon statistics of the retrieved field \cite{Bariani2012,Stan2012}, since the retrieval of states with more than one photon is suppressed.  This relies on the emission of multiple photon states being outside the phase-matched mode defined by the signal field.  It is important to note that the number of polaritons written into the sample is probabilistic (EIT and photon storage are coherent processes, so we would expect a superposition of polariton number states \cite{Bariani2012}; however, for simplicity we will assume a statistical mixture in this explanation because the two situations will give the same results in our experiments).  We operate in a regime where the medium is not saturated with Rydberg excitations.  The signal field is in a coherent state and therefore the number of polaritons should follow a Poisson distribution (which is truncated due to dipole blockade).  Dephasing can only occur when multiple polaritons are written into the sample.  In some experimental realisations only one polariton will be formed.  In these cases no interaction induced dephasing can occur, and a single photon should be retrieved.  If the dephasing process is effective in destroying the directionality of the read-out, there should be an excess of single photons retrieved relative to multi-photon retrievals.  As shown in figure \ref{Apparatus}(c), after applying a $2\pi$ microwave pulse the value of $g^{(2)}(0)$ decreases dramatically, with $g^{(2)}(0)=0.32 \pm 0.18$.  The enhanced anti-bunching of the retrieved field suggests that single Rydberg excitations are preferentially retrieved from the sample due to interaction-induced dephasing of the polaritons.  The microwave field therefore provides external control of the optical photon-photon interaction.  Optimisation of the anti-bunching should be possible through a more careful choice of the microwave pulse parameters, or using a more complex pulse sequence.  Note, though, that theory predicts $g^{(2)}(0)=0$ (complete dephasing of multiple Rydberg excitations) only in the limit of long dephasing times \cite{Bariani2012}.  The measurement of $g^{(2)}(0)$ after applying a $2\pi$ microwave pulse was only made possible subsequent to the publication of reference \cite{Maxwell2013}.  Interference filters were used to reduce the background light signal (i.e the signal obtained without any atoms), and in addition values of $g^{(2)}$ were background corrected using the measured signal to noise ratio \cite{Orrit2002}.  The extinction of the control field was also improved, leading to an increased storage signal. These changes meant that smaller retrieved signals could be measured.

For the $g^{(2)}$ measurements described above the entire retrieved pulse was binned, ignoring any variation within the pulse. However, since the polaritons must be spatially separated in the cloud by a distance $R \ge R_{\mathrm{o}}$ \cite{Gorsh2013}, one may expect this to be evident in the temporal photon correlations within each retrieved pulse \cite{Pohl2010}. To examine this, $g^{(2)}(0)$ is calculated for different regions of the retrieved pulse as shown in figure \ref{fockfil}(a). Note that no microwave coupling has been applied in this case. Examples of the analysis windows over which the retrieved pulse is studied are shown. All photon counts outside the highlighted analysis windows are ignored. The width of the windows is varied, starting from one edge of the pulse and moving towards the other edge. The windowing process is considered starting from both sides of the pulse, shown by the separate blue and red data points. It can be seen that the variation of $g^{(2)}(0)$ is strongly asymmetric with respect to the direction over which the analysis window is scanned. When fixing one end of the analysis window at the trailing edge of the retrieved pulse (red circles), the variation of $g^{(2)}(0)$ exhibits a step-like structure, decreasing quickly as the width of the analysis window is reduced. However, when fixing the analysis window at the leading edge of the pulse (blue squares), $g^{(2)}(0)$ remains roughly constant before increasing for small window widths. To gain some insight into the origin of the observed variation in $g^{(2)}(0)$ we consider the effect of dipole blockade on the group delay of the signal light. Due to the collective nature of the Rydberg excitations, modes with different photon numbers will experience different group delays. The dependence on photon number arises because the optical depth experienced by each photon depends on the number of photons in the medium. A single photon which collectively excites $N$ atoms experiences an optical depth, and therefore a group index, which is enhanced by a factor of $N$ compared to absorption by a single atom. However when two Rydberg polaritons are created they must be separated during propagation by a distance $R_{\rm{o}}$, and therefore each photon collectively excites a smaller number of atoms, of order $N/2$. Consequently the enhancement of the optical depth is reduced. Each mode of a propagating light field in a mixed state of photon numbers will propagate with a different group velocity. The observed variation in $g^{(2)}(0)$ across the retrieved pulse is consistent with modes which are temporally separated due to their different group velocities. 
%The resulting separation of the modes may be responsible for the observed variation in $g^{(2)}(0)$ across the retrieved pulse. 

It would be interesting to perform the same analysis shown in figure \ref{fockfil}(a) for the case where a microwave pulse is applied during the storage interval. Unfortunately this was not possible in the current experiment due to the lower photon count rates which naturally accompany the stronger suppression of $g^{(2)}(0)$. The detected photon count rate is reduced by around a factor of 3 after applying a $2\pi$ microwave pulse, giving around 0.003 photons per store-and-retrieve cycle on average (approximately 1 photon per second given the repetition rate of the experiment). As an alternative diagnostic, in figure \ref{fockfil}(b) we compare the shape of the retrieved pulse obtained with and without microwave coupling. Despite the fact that the pulse retrieved after microwave coupling has an amplitude which is roughly half that of the case where no microwave coupling is applied, the two pulses have a large degree of overlap in their trailing edges (purple highlighted region). It is the trailing edge of the pulse where the single-photon modes should dominate, since these should propagate most slowly through the medium. The data thus provides further evidence that the microwaves only cause dephasing when multiple polaritons are written into the sample, as the single-photon mode is not suppressed. In addition, the pulse shapes are consistent with the hypothesis that modes containing different photon numbers propagate at different speeds through the medium. This is highlighted in the inset of figure \ref{fockfil}(b), which shows the normalised difference between the two pulses. At short times, where modes containing multiple photons should dominate, the difference between the two pulses is greatest since the microwave field causes dephasing of these modes. At longer times the difference steadily decreases suggesting that the photon number of the contributing modes is also decreasing.  

\begin{figure}[!bt]
\includegraphics[width=\columnwidth]{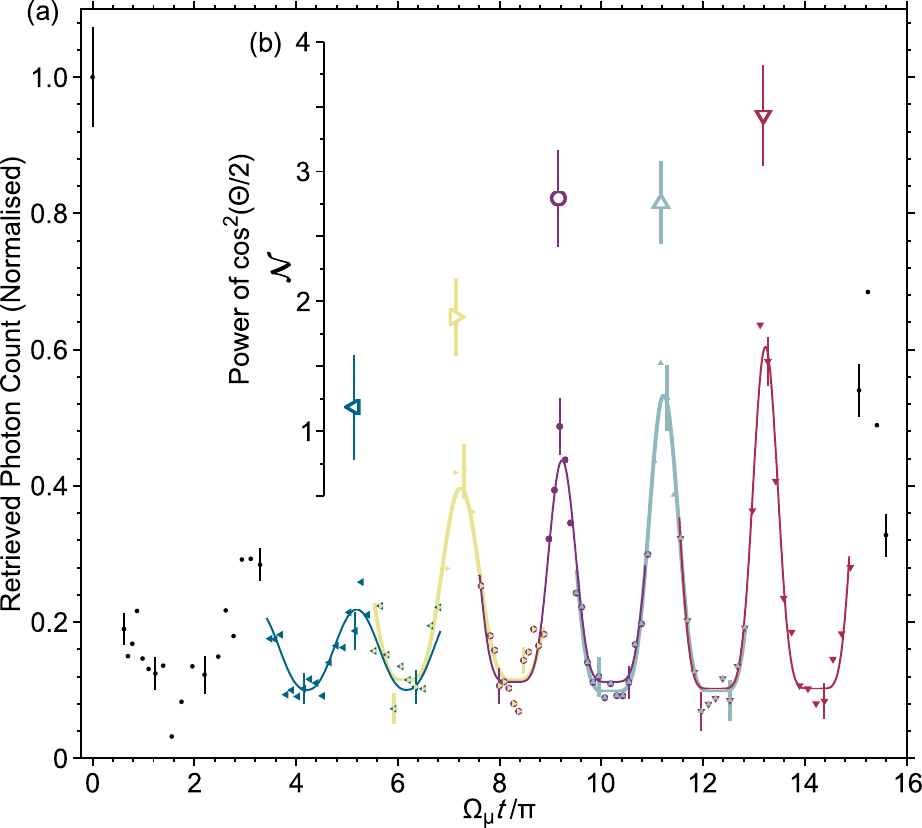}
\caption{(Color online)  Collective read-out of polariton fields containing different photon numbers. (a) Rabi oscillations of the collective $\cal{N}$-polariton state between microwave-coupled Rydberg levels are mapped out by varying the microwave Rabi frequency, $\Omega_{\mu}$, for fixed time $t=150$~ns. The retrieved photon signal (small filled symbols) is normalised to the case where no microwave coupling is applied. The function $P=[\cos^{2}(\Omega_{\mu}t/2)]^{\cal{N}}$ is fit to the data, where each peak of the oscillation is fit individually over a region defined by a sliding window.  The colour and shape of the small symbols indicate regions of data that are used together for a fit; some data points take part in more than one fit, shown by overlapped symbols.  The solid curves show the fits to the corresponding data regions, with all parameters (vertical offset, vertical scale, frequency, phase, and $\cal{N}$) independently fitted for each region.  A portion of these data were previously published as figure 3(b) in \cite{Maxwell2013}, though fitting with variable $\cal{N}$ is new to this paper.  (b, Inset) Large hollow symbols: The variation of the fit parameter $\cal{N}$ across the Rabi oscillations. In the regime where $\Omega_{\mu}<V_{\mathrm{dd}}$, $\cal{N}$ approaches one suggesting that only a single polariton survives.}
\label{NvsPeak}
\end{figure} 

The dephasing mechanism is also validated by the form of the Rabi oscillations of the polariton state between the microwave coupled levels. The Rabi oscillations, shown in figure \ref{NvsPeak}, have a strong many-body character. The microwaves excite Dicke states of $\cal{N}$ polaritons, resulting in Rabi oscillations in the retrieval probability of the form, $P=[\cos^{2}(\Omega_{\mu}t/2)]^{\cal{N}}$. The oscillations have this form since the $\cal{N}$-particle Rabi oscillations couple to a single optical read-out mode. Resonant dipole-dipole interactions during the period of microwave coupling can have a dramatic effect on the dynamics of the system. For strong microwave fields where $R_{\mu}<R_{\mathrm{o}}$, the dipole-dipole interactions are weak compared to the microwave driving. High amplitude Rabi oscillations can therefore be observed in this regime. However, for weak microwave fields where $R_{\mu}>R_{\mathrm{o}}$ the resonant dipole-dipole interactions dominate the microwave driving. Dephasing of the polaritons results in a suppression of the retrieved signal. To see whether the dephasing is reflected in the collective read-out of the polaritons, the value of $\cal{N}$ is studied for different ranges of $\Omega_{\mu}$. Each peak of the Rabi oscillations is fit individually over a range defined by a sliding window to extract a value of $\cal{N}$. It can be seen in the inset to figure \ref{NvsPeak} that, over the range considered, $\cal{N}$ increases with $\Omega_{\mu}$. The first Rabi oscillation gives a value of $\cal{N}$ close to 1, supporting the theory that only single polariton modes are retrieved in the strong dephasing regime. As $\Omega_{\mu}$ is increased the dephasing of the polaritons is reduced and $\cal{N}$ increases. The value of $\cal{N}$ obtained in the regime of strong microwave driving should therefore eventually saturate at a value which reflects the mean number of polaritons in the sample. This saturation is consistent with a regime where the time scale of the dephasing is slow compared to the microwave driving \cite{Bett2013}. 

In conclusion, we have built on our previous work \cite{Maxwell2013} by demonstrating that microwave-induced resonant dipole-dipole interactions destroy the phase-matched read-out of multiple photons from a cold atomic cloud.  Single-photon states however remain unaffected, as predicted theoretically (qualitatively) in reference \cite{Bariani2012} (unfortunately the quantitative predictions are not directly comparable, since they assume the initial storage is \emph{un}blockaded, and then use a microwave Ramsey sequence instead of a single pulse).  This provides a new tool of using microwave fields to assist the creation of controllable non-classical photon states, and is further evidence that Rydberg atoms are suitable for mediating photon-photon interactions in a multi-qubit quantum logic gate, and to provide an interface between the optical and microwave domains \cite{Petro2008}.

An open question is to whether the spectral properties of single photons produced using this method are suitable for their use in quantum logic gates.
% Currently we have no way to measure the spectral properties of the emitted photons, but an interesting direction for future work would be to examine the possibility of producing Fourier-limited single photons.
Analysis of the coherence of the retrieved photon signal, for example using homodyne detection \cite{Lvovsky2012, Grangier2014}, will be the focus of future work.
In addition, the observed dependence of the group delay on the photon number propagating inside the medium also has potential applications in the generation of Fock states.  Although the temporal separation of these states is limited in the current setup by the low optical depth of the medium (OD$\approx1$), much higher optical depths are possible \cite{Peyronel2012}.  An interesting direction for future work would be to correlate the observed photon statistics with spatially resolved measurements of the Rydberg polaritons \cite{Olmos2011,Gunt2012,Gunt2013}.  It may also be possible to tailor the shape of the photon modes so that they can be more clearly resolved \cite{Kuhn2010}.  

We thank C. Ates, S. Bettelli, T. Fernholz, and I. Lesanovsky for stimulating discussions.  We also thank
the referees for their useful comments.  The data in this paper are available on request.  This work was supported by Durham University; the Engineering and Physical Sciences Research Council grant reference EP/H002839/1; and the EU Marie Curie ITN COHERENCE grant reference FP7-PEOPLE-2010-ITN-265031.

\end{document}